\documentclass[%
reprint,
showkeys,
superscriptaddress,
showpacs,
nofootinbib,
 amsmath,amssymb,
 aps,
 prx,
 longbibliography,
floatfix,
]{revtex4-2}

\usepackage{refcount}
\usepackage{float}
\usepackage{stackengine}
\usepackage[separate-uncertainty=true]{siunitx}
\usepackage{wasysym}
\usepackage{mathrsfs}
\usepackage{graphicx}
\usepackage{dcolumn}
\usepackage{bm}
\newcommand{\vect}[1]{\boldsymbol{\mathbf{#1}}}

\usepackage[dvipsnames]{xcolor}
\usepackage[colorlinks,linkcolor=BlueViolet,citecolor=BlueViolet,urlcolor=BlueViolet]{hyperref}

\begin{document}

\preprint{}

\author{Alessandro Zunino}
\affiliation{Nanophysics, Istituto Italiano di Tecnologia, Via Morego 30, 16163 Genoa, Italy}
\affiliation{Physics Department, University of Genoa, Via Dodecaneso 33, 16146, Genoa, Italy}
\author{Salvatore Surdo}
\email{salvatore.surdo@iit.it}
\affiliation{Nanophysics, Istituto Italiano di Tecnologia, Via Morego 30, 16163 Genoa, Italy}
\author{Mart\'i Duocastella}
 \email{marti.duocastella@iit.it}
 \affiliation{Nanophysics, Istituto Italiano di Tecnologia, Via Morego 30, 16163 Genoa, Italy}

\title{Dynamic multi-focus laser writing with acousto-optofluidics}

\begin{abstract}
Laser writing of materials is normally performed by the sequential scanning of a single focused beam across a sample. This process is time-consuming and it can severely limit the throughput of laser systems in key applications such as surgery, microelectronics, or manufacturing. Here we report a parallelization strategy based on ultrasound waves in a liquid to diffract light into multiple beamlets. Adjusting amplitude, frequency, or phase of ultrasound allows tunable multi-focus distributions with sub-microsecond control. When combined with sample translation, the dynamic splitting of light leads to high-throughput laser processing, as demonstrated by locally modifying the morphological and wettability properties of metals, polymers, and ceramics. The results illustrate how acousto-optofluidic systems are universal tools for fast multi-focus generation, with potential impact in fields such as imaging or optical trapping. 
\end{abstract} 


\maketitle

\section{Introduction}

Controlled delivery of light at targeted positions on a sample is key for precise, direct, and localized modification of materials using laser-based systems \cite{Chong2010, Duocastella2012}. Typically, laser writing is performed by focusing a beam into a single spot, which is successively scanned across a region of interest by either mirrors or translation stages \cite{Osellame2011, Pique2016}. Even if sub-wavelength control of light can be achieved by tuning focusing-optics and scanning \cite{Sahin2014}, the sequential nature of this approach seriously constrains processing throughput. A promising way to address this issue is by splitting the incident beam into several beamlets, each focused at a different position on the sample. In this case, throughput increases linearly with the number of beamlets --- up to five orders of magnitude improvement has been demonstrated \cite{Eggeling2013} --- only limited by the available laser power. Successful implementations include the use of passive elements to generate fixed multi-focus distributions, such as diffractive optical elements \cite{Kuroiwa2004}, amplitude gratings \cite{Nakata2004} or beam splitters \cite{Cheng2011}. However, the lack of tunability in the selection of focus location or number heavily limits the flexibility of laser processing. Alternatively, active elements such as acousto-optic deflectors (AODs) \cite{Dugan1997, Hafner2018}, spatial light modulators (SLMs) \cite{Obata2010, Sun2018} , or digital micromirror devices (DMDs) \cite{An2014, Geng2019} enable dynamic and customized light splitting, but can suffer from polarization dependence, aberrations, long response time, low damage threshold or pixelization issues. Simply put, a parallelization strategy for multi-focus generation that grants high tunability, low aberrations, and speed is still missing.

Here, we exploit the interaction between light and sound in a liquid to create dynamic multi-focus distributions with unprecedented tunability and speed. Our core idea is to use mechanical vibrations in a liquid to induce harmonic variations of the liquid density, and hence its refractive index. As a result, a laser beam interacting with such inhomogeneous media is diffracted into multiple beamlets (figure \hyperref[fig:setup]{\ref{fig:setup}a}), whose number or intensity can be selected based on the frequency or amplitude of the acoustically-induced density changes. Various acousto-optofluidic (AOF) systems have been successfully employed for rapid light focusing in microscopy \cite{Piazza2018, Kong2015} and laser material processing \cite{Duocastella2013, Chen2018}. Recently, they have been used to generate interference patterns in maskless lithography \cite{Surdo2019}. Nonetheless, this is the first time, to the best of our knowledge, that an AOF system is used for customized multi-focus generation in laser material processing. We present a detailed optical characterization of our system along with a full analytical treatment that is in perfect agreement with experiments. By using different processing modes, namely discrete, continuous, and brush-like scan, we demonstrate tailored micromachining of metals, ceramics, and polymers at ease and high-throughput. As a proof of concept, we fabricated functional surfaces with locally tuned wettability, down to the sub-micrometer scale. 

\section{Dynamic multi-focus generation}

\begin{figure}
\centering
\includegraphics[width=1\columnwidth]{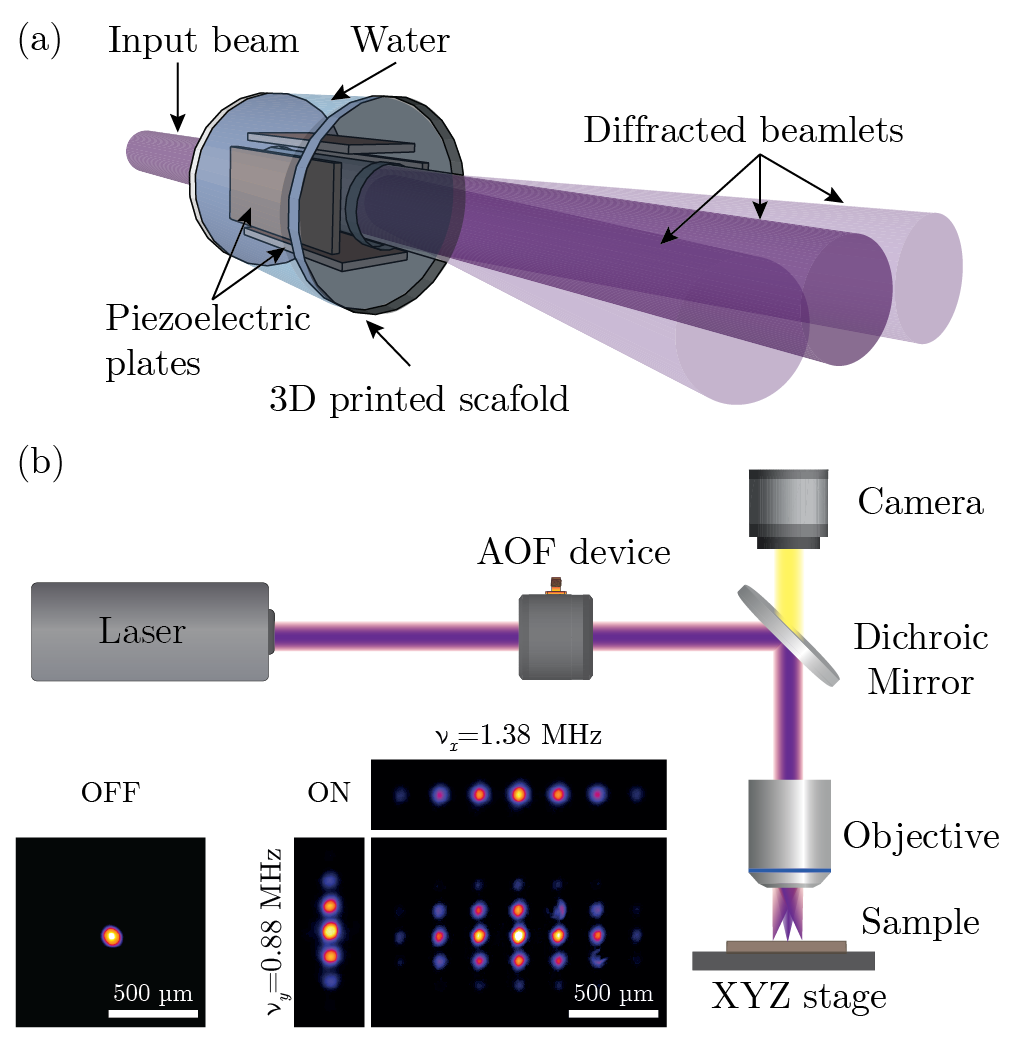}
\caption{Laser multi-focus generation with acousto-optofluidics (AOF). (a) Working principle of the AOF device. A laser beam passes through the acoustic cavity, where oscillating piezoelectric plates maintain standing mechanical waves that diffract light into multiple beamlets (Raman-Nath diffraction). (b) Scheme of the AOF-enabled laser direct-writing system. The inset corresponds to the measured intensity of a focused beam with the AOF off (bottom-left) and AOF on (bottom-right). For this latter, the device was driven at two different frequencies along the $x$-axis and $y$-axis. Note that the multi-focus distribution when driven with both axis simultaneously is not simply the sum of each individual axis.}
\label{fig:setup}
\end{figure}

A scheme of our laser multi-focus writing system is shown in figure \hyperref[fig:setup]{\ref{fig:setup}b}. Briefly, we modified a conventional Laser Direct Writing (LDW) system to include the acousto-optofludic device at a conjugate plane of the back pupil of the focusing objective. The AOF device consists of two pairs of piezoelectric plates aligned along orthogonal directions and placed in a sealed cylindrical cavity enclosed with two optically transparent windows. In experiments herein, the cavity was filled with deionized water, but any weakly absorbing liquid could be used as well. The choice of water is particularly convenient, being an isotropic and homogeneous medium.

This provides a polarization-insensitive system, in contrast to typical acousto-optic devices usually built with birefringent crystals. By driving  the piezoelectric plates at cavity resonance conditions, acoustic standing waves are established in the liquid, thus inducing temporal gradients in refractive index that enable controlled diffraction of light. Specifically, we operated our device at conditions where the dominant acousto-optic interaction was Raman-Nath diffraction \cite{Ohtsuka1985}(see Appendix). In this case, the periodic modulation of refractive index can be modeled as a phase grating whose far-field diffraction pattern consists of multiple diffraction foci. In particular, if the input is a normally incident Gaussian beam with peak intensity $I_0$, and only a pair of piezoelectic plates is driven, the intensity at the focal plane of the objective lens calculated with classical wave optics is (see Appendix)
\begin{multline}
    I(x,y,t)=I_0\sum_{q=-\infty}^{+\infty}J^2_q\big[\kappa ln_m\cos(\omega_mt)\big] \times \\
    \times \exp\left\{-\frac{2}{w^2}\left[\left(x-\dfrac{qk_m\lambda f}{2\pi}\right)^2+y^2\right]\right\},
    \label{eq:intensity}
\end{multline}

\noindent where $q$ is the diffraction order, $J_q$ is the Bessel function of the first kind, $\lambda$ is the wavelength of the input beam, $\kappa=2\pi/\lambda$ is the wave number, $f$ is the focal length of the objective lens placed after the device, $w$ is the waist radius of the diffraction beam, $L$ is the longitudinal length of the AOF device,$c_s$ is the speed of sound in the liquid, $\omega_m=c_sk_m$ is the $m^{\text{th}}$ resonant angular frequency of the cavity, and $n_{m}$ is the induced change in refractive index (which depends on the driving voltage $V$ as $n_m \propto V/\omega_m$; see Appendix). Notably, equation \ref{eq:intensity} corresponds to a comb of Gaussian beams, each with an efficiency $\eta_{q}$, namely the peak intensity of the $q$-th diffraction order normalized with respect to the undiffracted beam peak intensity, given by
\begin{equation}
    \eta_q(t) = J^2_q\big[\kappa ln_m\cos(\omega_mt)\big]
    \label{eq:eta_q}
\end{equation}

\noindent and separated by a distance of

\begin{equation}
    \Delta\mu =\frac{\omega_m}{2\pi c_s}\lambda f.
    \label{eq:separation}
\end{equation}

\noindent Equations \ref{eq:eta_q} and \ref{eq:separation} describe the fundamental operating principle of our multi-focus generator --- by adjusting driving frequency, amplitude voltage or synchronized laser arrival time, the number, intensity or separation of the multiple diffraction orders can be tuned. Considering the superposition principle of acoustic waves, these equations could be generalized to include the contributions of multiple resonant frequencies (see Appendix).

\section{Temporal characteristics and tunability}

\begin{figure*}[tb]
\centering
\includegraphics[width=1\textwidth]{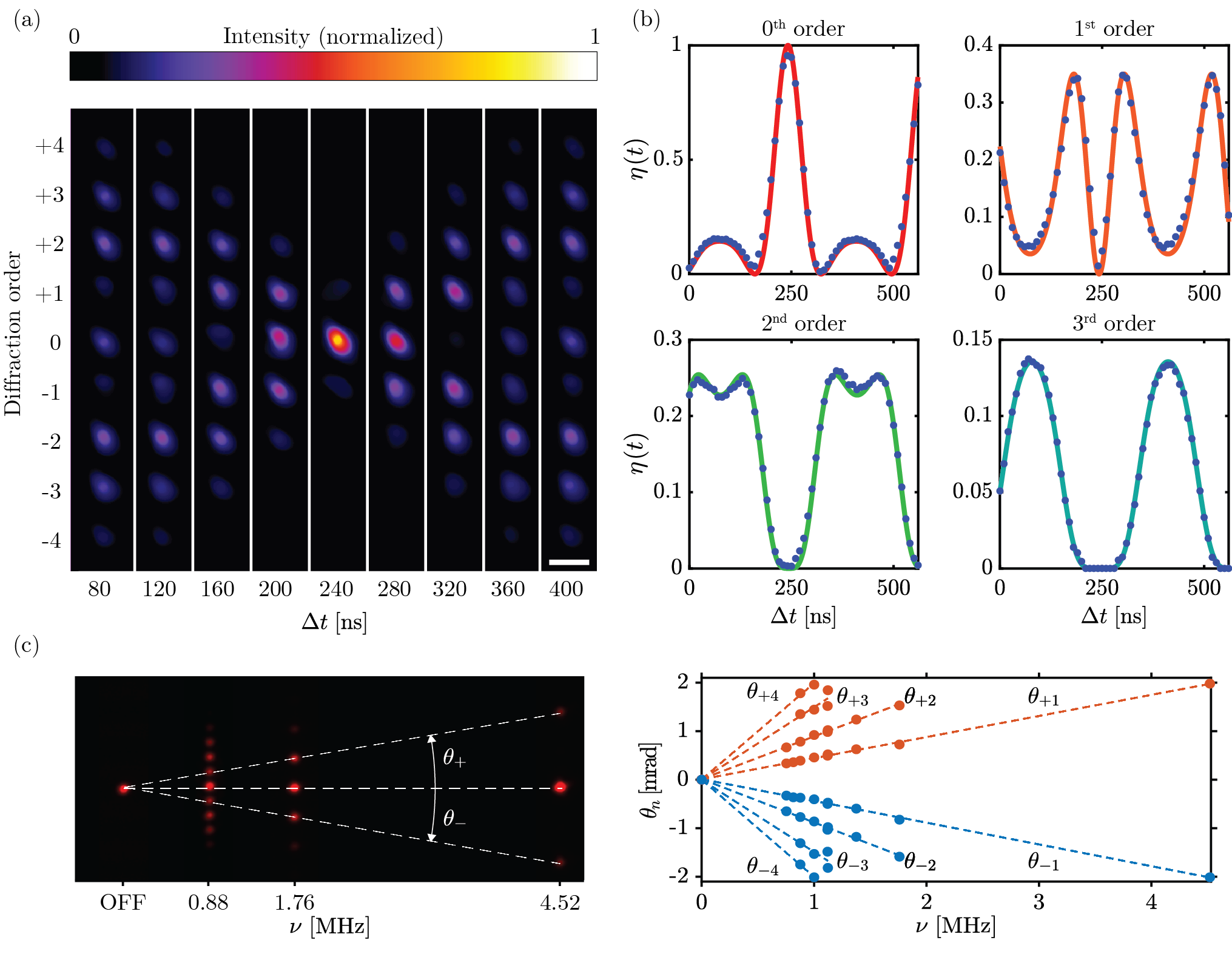}
\caption{Temporal characteristics of the AOF multi-focus generator. (a) Images of multi-focus distributions acquired with synchronized laser pulses at $V_{pp}=\SI{20}{V}$ and $\nu=\SI{1.488}{MHz}$. The pulse duration was $\SI{5}{ns}$ while the delay between consecutive measurements was \SI{10}{ns}. Scale bar \SI{100}{\micro m}. (b) Corresponding diffraction efficiency (blue dots) over time for the first four orders. Solid lines are fits to Equation \ref{eq:eta_q}. (c) Images of multi-focus distributions generated with a CW laser ($\lambda=\SI{645}{nm}$) for various driving frequencies (left), and corresponding diffraction angles (right). Dashed lines are linear fits to Equation \ref{eq:angle}.}
\label{fig:eta-fit}
\end{figure*}

 To study the speed at which the AOF system can switch between different multi-focus distributions, we recorded the diffraction foci with a camera placed at the focal plane of a converging lens. In detail, we used synchronized laser pulses with a duration much shorter than the AOF driving period and acquired snapshots of the light interaction with the instantaneous density oscillations induced in the liquid. As shown in Figure \hyperref[fig:eta-fit]{\ref{fig:eta-fit}a}, changing the time delay $\Delta t$ between pulses and the acoustic standing wave results in different distributions of Gaussian spots. Notably, the individual intensity of various diffraction orders --–  up to nine in this experiment --– can be adjusted in a time scale of tens of nanoseconds. Such remarkable speed is well above the limits of traditional AODs, which use traveling acoustic waves and have a typical response time of microseconds, mainly limited by clear aperture and speed of sound \cite{Maydan1970, Johnson1979}. 

\begin{figure*}[tb]
\centering
\includegraphics[width=1\textwidth]{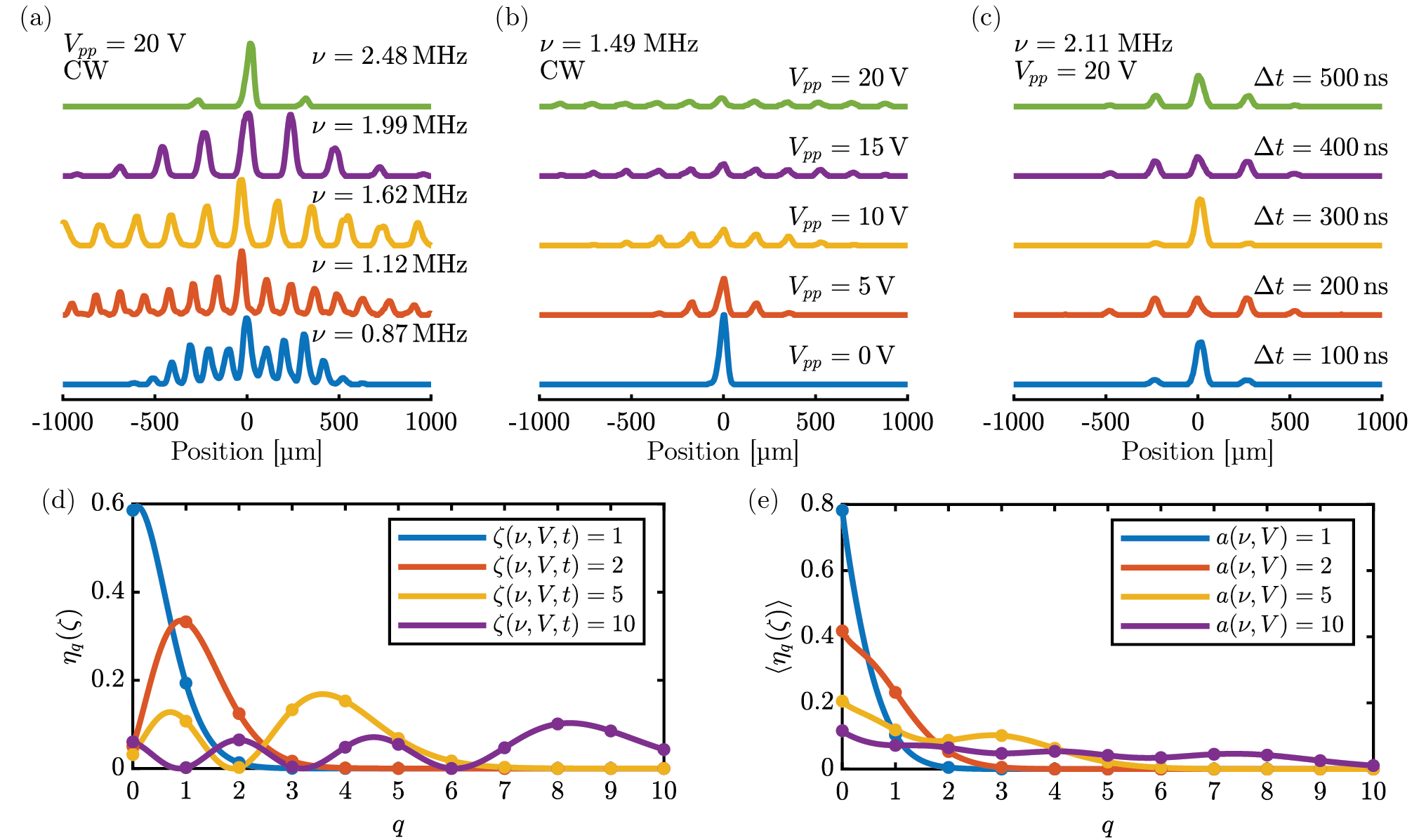}
\caption{Experimental and computational study of the diffraction efficiency as a function of the main driving parameters: peak-to-peak Voltage $V_{pp}$, resonant frequency $\nu$ and time delay $\Delta t$. (a) Intensity profiles at $V_{pp}=\SI{20}{V}$ across the peaks of multi-focus distributions obtained at different driving frequencies and using a CW laser as the illumination source. (b) Intensity profiles at $\nu=\SI{1.49}{MHz}$, using a CW laser, and for different driving voltage amplitudes. (c) Intensities measured at $\nu=\SI{2.11}{MHz}$, $V_{pp}=\SI{20}{V}$, in synchronized mode, and at various time delays. All the experiments were performed using a blue laser ($\lambda=\SI{445}{nm}$) as light source. (d) Simulated instantaneous diffraction efficiency for various arguments $\zeta=a\cos(\omega t)$ and diffraction orders $q$. (e) Simulated averaged diffraction efficiency with fixed $a=\kappa L n_m$ (the integral mean of equation \ref{eq:eta_q} over a period has been calculated numerically). In both plots the functions are presented as continuous just to guide the eye, but the physical relevant case is only when $q$ is an integer.}
\label{fig:parameters}
\end{figure*}

The high-speed capabilities of the AOF system originate from the continuous operation at steady-state --– we only adjust the synchronization delay, which is solely limited by electronics. However, access to additional multi-focus distributions is possible by tuning frequency or amplitude of the driving signals. In this case, there is an inherent transition time until the acoustic standing wave reaches a new steady-state. For water, it has been measured to be about \SI{600}{\micro s}, which corresponds to a frequency of approximately \SI{1.7}{kHz} \cite{Surdo2019}. While lower than before, this speed it is still significantly higher than most state-of-the-art active multi-focus generators. Interestingly, this is also the maximum speed attainable with continuous-wave (CW) lasers or laser pulses longer than the AOF driving period. Indeed, the loss of any temporal dependency of the instantaneous refractive index --- only an average acousto-optic interaction is detected in these instances --- makes driving frequency or voltage the only parameters available for modifying the diffraction pattern.

Generating multi-focus distributions with AOF is not only fast, but also predictable. Figure \hyperref[fig:eta-fit]{\ref{fig:eta-fit}b} displays the temporal evolution of the diffraction efficiency $\eta_{q}$ for a single resonant frequency and four different values of $q$. Experimental results are in perfect agreement with the theoretical behavior described by equation \ref{eq:eta_q}. Similarly, adjusting the driving frequency produces multi-focus distributions at predictable spatial coordinates. As shown in Figure \hyperref[fig:eta-fit]{\ref{fig:eta-fit}c}, the angular spreading of the diffraction orders is symmetric and increases linearly with frequency. This behavior is in agreement with Equation \ref{eq:separation}, from which the diffraction angle for each order can be written as 

\begin{equation}
    \theta_q =\arctan \left(q\frac{\omega_m\lambda}{2\pi c_s}\right)\sim q\left(\frac{\nu_m\lambda}{c_s}\right),
    \label{eq:angle}
\end{equation}
where linearization is possible because the acoustic wavelengths $c_s/\nu_m$ are much larger than $\lambda$. Therefore, for a given focal length, the location of the diffraction orders can be precisely selected by tuning resonant frequency. In practice, there exists a frequency limit below which the separation between orders is too small to be resolved, namely if $\Delta \mu\lesssim w$. Still, this scenario could be of interest for generating large spots with a customized intensity distribution.

A more in-depth experimental analysis of the different multi-focus distributions accessible when varying the main parameters, namely driving frequency, amplitude, and delay time is shown in Figure \ref{fig:parameters}. As expected from theory, the separation between diffraction peaks increases with driving frequency (Figure \hyperref[fig:parameters]{\ref{fig:parameters}a}). Interestingly, the driving frequency also controls the diffraction efficiency of high order modes --– the energy carried by higher modes decreases with this parameter. This can be explained considering the inverse relationship between the changes in refractive index induced in the liquid and the driving frequency (see Appendix). Importantly, this effect can be compensated by increasing the amplitude $V$ of the driving signal that results in a larger number of diffraction orders (Figure \hyperref[fig:parameters]{\ref{fig:parameters}b}). Note, though, that as more orders are present, conservation of energy causes the efficiency of every order to decrease. Such a trend can be broken by using synchronized pulsed illumination, as shown in Figure \hyperref[fig:parameters]{\ref{fig:parameters}c}. In this case, the time delay acts as an additional degree of freedom, and by adjusting this parameter the efficiency of user-selected diffraction orders can be tuned. 

The role of the different technological parameters can be properly described by using equation \ref{eq:eta_q}. Indeed, all the complex multi-focus distributions experimentally observed can be predicted with a single variable, $\zeta$, defined as the argument of the diffraction efficiency $\eta_q$: 

\begin{equation}
    \zeta (\nu,V,t)=a(\nu,V)\cos(\omega_mt)=\kappa L n_m \cos(\omega_mt).
    \label{eq:zeta}
\end{equation}
Such functional dependency is graphically shown in Figure \hyperref[fig:parameters]{\ref{fig:parameters}d-e}. In all cases, a maximum diffraction order with a non-zero efficiency exists that depends on the driving voltage amplitude. For synchronized pulsed illumination, the efficiency of each individual order can be tailored, enabling higher diffraction orders to carry more energy than lower ones, even the zeroth. Instead, for CW illumination or when using laser pulses with a duration longer than the period of the driving oscillation, only an averaged efficiency can be attained. As a result, the diffraction efficiency decreases with the diffraction order.

\section{Multi-focus processing of materials}

\begin{figure*}[htb]
\centering
\includegraphics[width=1\textwidth]{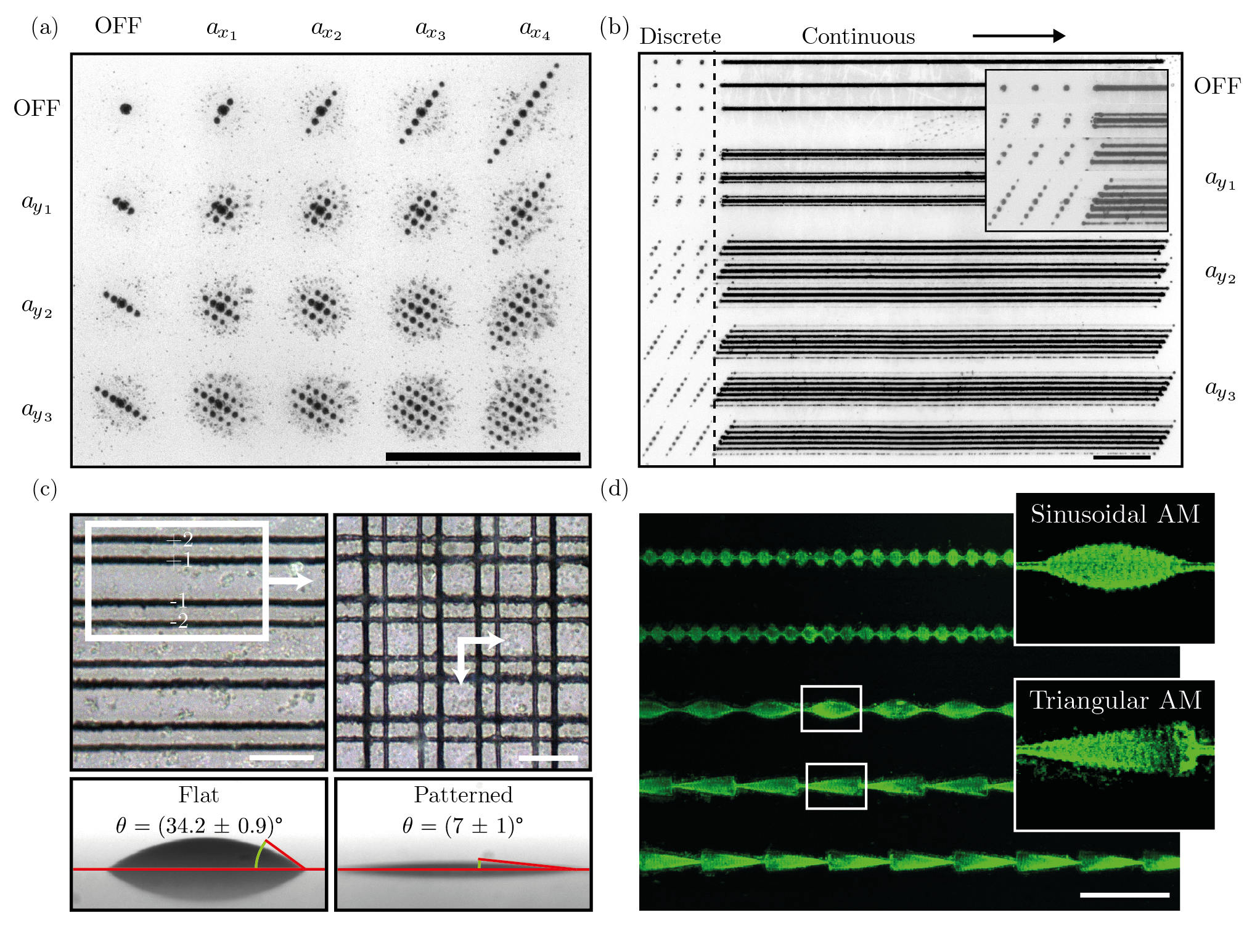}
\caption{Materials processing with AOF-enabled LDW workstation. (a) Result of multi-focus laser ablation of chromium for different driving conditions. For each multi-focus distribution, 25 asynchronous laser pulses were fired. (b) Direct writing of lines on chromium with the AOF device on and the stage continuously moving. The stage speed was selected so that, at each position, multiple asynchronous laser pulses were shot. (c) Top-left: ablation of borosilicate glass using the AOF device with synchronized laser pulses. Note that the $0^\text{th}$ order is missing, while the orders $\pm 1$ and $\pm 2$ generated lines of similar width. Top-right: scanning along the orthogonal direction resulted in a 2D grid. Bottom: contact angle measurements of water on a glass substrate before (left) and after laser patterning (right). (d) Wide-field fluorescence image of PDMS-coated glass after irradiation with the laser operated in brush-like mode and after wetting with an aqueous fluorescein solution. Only the irradiated parts are hydrophilic and therefore fluorescent. White scale bars \SI{10}{\micro m}. Black scale bars \SI{50}{\micro m}.}
\label{fig:application}

\end{figure*}

The simple integration of the AOF device in a laser direct-writing workstation enables modification of materials at a high throughput. Figure \hyperref[fig:application]{\ref{fig:application}a} shows different ablated spots on a chromium thin film obtained by irradiation with the AOF-enabled system. Size, distance or quantity of the ablated spots can be easily tuned by adjusting the driving conditions of the AOF device without need of any mechanical motion of the sample. This represents a key advantage of our approach -the throughput of an LDW system can be increased by a factor that scales linearly with the number of diffracted foci.

Combining the tunability offered by acousto-optofluidics with sample translation enables writing complex structures (Figure \hyperref[fig:application]{\ref{fig:application}b-d}, Supplementary Movie 1, Supplementary Movie 2). Notably, different processing modes can be distinguished based on the type of illumination or stage speed. For CW lasers (or multiple asynchronous laser pulses) and a speed $v > d\nu$ (where $\nu$ is the driving frequency of the acoustic cavity and $d$ is the diameter of the ablated spot), continuous lines can be generated (Figure \hyperref[fig:application]{\ref{fig:application}b}). Because of the higher efficiency of the low diffraction orders, the lines become progressively thinner away from the center. Instead, operating with synchronized laser pulses enables generating structures in which the line width of each diffraction order can be tuned. The structures can be rendered continuous or discrete by simply adjusting the stage speed relative to the laser repetition rate. As a proof of concept, we used this patterning mode to locally modify the wetting properties of a target surface (\hyperref[fig:application]{\ref{fig:application}c}). By only selecting the first and second diffraction orders, and by scanning the multi-focus distribution along the $x$ and $y$ axis, we obtained a grid-like pattern on glass. As expected from Wenzel's model \cite{Wenzel1936}, the patterned surface exhibited an increase in roughness, and consequently, an enhancement of its wetting behavior ---the patterned glass was transformed into a super-hydrophilic surface.

The capability of the AOF-enabled workstation for generating functionalized structures is shown in Figure (\hyperref[fig:application]{\ref{fig:application}d}). In this case, we used a laser to locally remove a hydrophobic polymeric coating (polydimethylsiloxane or PDMS) \cite{Eduok2017} from a glass substrate. As a result, only the laser irradiated areas became hydrophilic. By driving the AOF device with a low driving frequency and by using a high laser fluence and asynchronous laser pulses, the different diffraction orders were merged. In other words, the laser beam acted as a brush, which thickness could be tuned by modulating the driving voltage amplitude. Therefore, laser beam scanning along one axis combined with sinusoidal or triangular modulation of the signal amplitude, allows fabricating significantly complex patterns at ease and high-speed (Supplementary Movie 3). Note that such brush-like mode can lead to virtually any arbitrary pattern, with the only constraint to be symmetrical with respect to the scanning axis. After dropping a fluorescein-containing aqueous solution on the patterned surface and waiting for evaporation to occur, the domains exhibiting different wettability can be distinguished. As expected, only the laser-irradiated areas were fluorescent, proving the feasibility of AOF systems for writing complex and functional patterns.

\section{Conclusion}

The variations in refractive index induced by ultrasound waves in a liquid-filled resonant cavity can be used to parallelize laser writing. Adjusting the sound parameters, namely frequency, amplitude and phase, results in user-selectable multi-focus distributions at high speeds, down to sub-microsecond time scales. This adds to the collection of degrees of control typically available in laser processing, thus providing, in a single setup, an unrivaled combination of throughput, speed and ease of use.  

We anticipate that acousto-optofluidics will help to overcome the intrinsic trade-off between flexibility and throughput of laser-direct writing systems. As our results demonstrate, driving the cavity at moderate voltage amplitudes (below \SI{20}{V}) complex patterns can be rapidly ablated on a surface. The benefits of our innovation could be similarly extended to laser additive manufacturing, including Laser Induced Forward Transfer (LIFT) \cite{Surdo2017}, laser sintering \cite{Hong2013} or multi-photon polymerization \cite{Fischer2013}. Outside the manufacturing field, precise and fast light splitting can also have a key role in the realization of optical traps \cite{Trypogeorgos2013} or in beam multiplexing for fast microscopy \cite{Cheng2011}. Furthermore, by using higher driving amplitudes, novel acousto-optic interaction regimes could be explored, such as chaotic or subharmonic \cite{Cantrell2015}, paving the way to shaping the light with unprecedented spatiotemporal control. 

\section{Materials and methods}
\label{sec:setup}
\paragraph*{Acusto-Optofluidic device:}

 The AOF device consisted of a hollow metallic cylinder ($\diameter=\SI{2.54}{cm}$), a couple of optically transparent windows (Thorlabs, WG41010), and a plastic scaffold for properly holding four piezoelectric plates (APC international, PZT Navy type II)with sizes of $\SI{20 x 9.5 x 1.5}{mm}$. We employed a 3D-printer (CubePro, 3D Systems with a resolution of \SI{100}{\micro m}) to build the scaffold using poly-lactic acid (PLA). The piezoelectric plates were soldered on each face with metallic wires and subsequently assembled together with the scaffold to form a cavity with square symmetry ---each plate corresponds to a face of a rectangular parallelepiped, and the distance between parallel plates is about $\SI{10}{mm}$. Note that each pair of piezoelectric plates can be independently controlled. The cavity was positioned inside the metallic cylinder, enclosed with a pair of optically transparent windows and sealed with an epoxy resin (Loctite, Hysol-9483). In experiments herein, the cavity was filled with Milli-Q water, and the AOF device was driven by an arbitrary waveform generator (Agilent 33521A).

\paragraph*{Optical characterization setup:}

The optical characterization of the AOF device was carried out by recording the multi-focus distribution with a CMOS camera (ThorLabs, DCC1545M). Specifically, a converging lens ($f=\SI{400}{mm}$) was used to generate the far-field diffraction pattern that was indeed imaged at the lens focus. The light source was a \SI{445}{nm} laser (Coherent CUBE 445-40C) that can be operated in either CW or pulsed mode. In the latter case, the laser was  externally triggered and a pulse delay generator (Stanford Research System, DG645) was used to control the delay time between laser pulses and the driving signal of the AOF device. This allowed us to send laser pulses through the cavity at specific phase values of the driving signal.

\paragraph*{Material processing setup:}

The laser direct-writing setup consisted of a Ti:Sapphire amplifier (Coherent Legend, pulse duration $=\SI{70}{fs}$, repetition rate $=\SI{1}{kHz}$, $\lambda=\SI{800}{nm}$), optics to guide and focus the beam into the sample, and a fast XYZ stage (Prior Scientific, Inc.). The last focusing element of the LDW workstation was a microscope objective (Mitutoyo M Plan Apo SL $50\times$,\SI{0.75}{NA}), and the AOF device was placed at a conjugate plane of the back focal plane of the objective using a 4f system. We also integrated an upright bright-field microscope into the LDW system for direct inspection of the material modification.
We controlled laser firing and stage motion by a computer using a custom software written in LabView. To perform material processing in synchronized mode, we used the laser as master, and triggered the AOF device using the voltage generator in burst mode. In this way, the sound waves were generated after a controllable delay time, chosen bigger than the time required for the cavity to reach steady state.

\paragraph*{Sample preparation and characterization:}
Metal surfaces were prepared by sputtering \SI{200}{nm} of Chromium on top of a glass substrate. The PDMS-coated glasses were obtained by \textit{dip coating}. Specifically, a microscope glass was quickly rinsed into a solution of PDMS in heptane (5\% v/v) resulting in a polymeric film with a thickness of approximately \SI{500}{nm}.
Morphological characterization of the ablated surfaces was performed with a bright-field optical microscope (DM2500 M, Leica). In particular, a $10\times$/\SI{0.25}{NA} and a $50\times$/\SI{0.75}{NA} microscope objectives were used for imaging the ablated patterns. Large field of view images of deposited fluorescein were recorded  with a wide-field epifluorescence microscope (Nikon eclipse 80i) equipped with a $10\times$/\SI{0.3}{NA} objectives (Plan Fluor, Nikon). Magnified images of the same samples were instead acquired with a $40\times$/\SI{0.75}{NA} microscope objective. 
The wetting behavior of the laser-irradiated surfaces was studied with an optical contact angle goniometer (Dataphyisics OCA 15EC) by depositing \SI{0.2}{\micro L} of deionized water on top of the sample under test. To ensure good reliability, 5 measurements were performed for each surface.

\appendix
\setcounter{equation}{0}
\renewcommand\theequation{A\arabic{equation}}
\section*{Appendix}

\subsection{Solution of the wave equation in the acoustic cavity}

Propagation of acoustic waves in a viscous medium, water in our case, is governed by the damped acoustic wave equation \cite{McLeod2007, Surdo2019}, which for a single axis can be written as

\begin{equation}
    \frac{\partial^2 \rho}{\partial t^2}- \frac{\partial^2}{\partial x^2}\left( \nu \frac{\partial \rho}{\partial t}+c_s^2\rho \right)=0,
    \label{eq:wave}
\end{equation}

\noindent where $\nu$ is the kinematic viscosity of water, $c_s$ is the speed of sound in water and $\rho$ is the variation of density.
To describe the effect of the oscillating plates, we applied the following boundary conditions:
\begin{subequations}
\begin{align}
    v|_{x=0}&=v_a\cos(\omega t) \\
    v|_{x=L}&=v_a\cos(\omega t)
\end{align}
\end{subequations}
where $v$ is the local fluid velocity, $v_a$ is the oscillation amplitude of the velocity, $L$ is the length of the cavity and $\omega$ is the angular frequency of the driving signal. In the resonant and low viscosity case, the solution of equation \ref{eq:wave} can be written as 

\begin{equation}
    \rho_m(x,t)= \frac{4\rho_0c_s^2d_{33}V}{\nu L\omega_m}\cos(\omega_m t)\cos\left(k_mx \right),
    \label{eq:density}
\end{equation}
where $k_m=\dfrac{\omega_m}{c_s}=\dfrac{m\pi}{L}\;(m\in 2\mathbb{N}+1)$ is the wave number, $\omega_m$ is the $m^{\text{th}}$ resonant angular frequency of the cavity, $V$ is the voltage amplitude applied to the piezoelectric plates and $d_{\alpha\beta}$ is the piezoelectric charge tensor \cite{Surdo2019}. In the above formula it has been assumed that $v_a=\omega_md_{33}V$.

Assuming the polarizability of water to be constant, we can use a linearized form of the Lorentz-Lorenz equation \cite{McLeod2007} to convert the variations of density into variations of refractive index:

\begin{equation}
    n(\rho_m)=n_0+\left.\frac{\partial n}{\partial \rho}\right|_{\rho_0}\!\!\!\!\rho_m=n_0+\frac{n_0^4+n_0^2-2}{6n_0}\frac{\rho_m}{\rho_0},
    \label{eq:lorentz-lorenz}
\end{equation}
where $n_0$ and $\rho_0$ are, respectively, the static refractive index and density of water.

Combining equations \ref{eq:density} and \ref{eq:lorentz-lorenz} it is possible to obtain an analytic expression for the instantaneous  refractive index in the cavity:   

\begin{equation}
    n(x,y,t)=n_0+\underbrace{\overbrace{\frac{4c_s^2d_{33}V}{\nu L\omega_m}\frac{n_0^4+n_0^2-2}{6n_0}}^{n_m}\cos\left( k_mx \right)\cos(\omega_m t)}_{\Delta n(x,y,t)}
\end{equation}

\subsection{Acousto-optic interaction regime}

To determine the type of diffraction that occurs when a beam passes through the AOF device, we can use the Klein-Cook parameter defined as \cite{Klein1967}:

\begin{equation}
    Q=\frac{2\pi\lambda l}{n_0\Lambda^2}
\end{equation}

\noindent being $\lambda$ the wavelength of the laser beam, $l$ the length of the piezoelectric plate ($\approx\SI{2}{cm}$ in current design), and $\Lambda$ the wavelength of the acoustic wave. Considering visible light, and a typical driving frequency of $\SI{1}{MHz}$ (with a corresponding wavelength of the acoustic wave of about $\SI{1.5}{mm}$), in our experiments $Q\approx0.03$. At this condition ($Q\ll 1$) the coupling between the different modes is highly efficient, resulting in multiple diffraction orders. It is therefore legit to use the Raman-Nath model \cite{Raman1} to calculate the Fraunhofer diffraction.

\subsection{Calculation of the diffraction intensity}

The equation of a Gaussian beam traveling along the $z$-axis is given by

\begin{multline}
    E(r,z) = E_0 \frac{w_0}{w(z)} \exp \left[ \frac{-(\vect{r}-\vect{r}_0)^2}{w^2(z)}\right] \times \\
    \times \exp \left[ -i\left(\kappa z +\kappa  \frac{r^2}{2R(z)} - \psi(z)\right) \right]
    \label{eq:gaussian}
\end{multline}
\noindent where $w_0$ is the beam waist, $w$ the beam width, $\kappa$ is the wave number, $R$ is the radius of curvature, and $\psi$ is the Guoy phase factor. 
The optical path length travelled by such a beam after passing through an acoustic cavity (with the beam propagation direction being orthogonal to the sound waves) is $\delta(x,y,t)=ln_0+l\Delta n(x,y,t)$. $ln_0$ is an additive constant phase factor and can be neglected for calculating the intensity. Using the numerical values of the typical experiments described in the manuscript, $ln_m$ is expected to have a value of the order of \SI{e-7}{m}: since this is the same order of magnitude of $\lambda$ we can neglect the contribution of the slowly varying (along the optical axis $z$) radius of curvature and Guoy phase factors \cite{Svelto2009} and just consider the propagation phase shift\footnote{Notice that if we use a well collimated beam (i.e. if $z\ll z_R$ and $z_R\gg l$) we can obtain the same results further described even if $\kappa ln_m\gg\lambda$. The only correction to perform is to remove all the terms containing the radius of curvature, because $R(z)\xrightarrow[z\ll z_R]{}+\infty$.}.
Therefore, we can describe the optical cavity with the following complex transmittance function
\begin{equation}
    t_A=\exp\big[i\kappa l\Delta n(x,y,t)\big]
\end{equation}
In order to simplify the next calculations, it is useful to rewrite the transmittance function  $t_A$ for a reference frame that is coaxial with the device:

\begin{equation}
    t_A= \exp\left[i\kappa ln_m\cos(\omega_mt)\sin\left(k_mx+\frac{m+1}{2}\pi\right)\right].
    \label{eq:trasmittance}
\end{equation}
Now we can write the scalar field\footnote{Here we can use the scalar diffraction theory because the frequency used in the experiments are at maximum of the order of \SI{10}{MHz} and therefore the wavelength of the acoustic wave is much bigger than the one of the light \cite{Goodman2017}.} emerging from the acoustic cavity as the product of equations \ref{eq:gaussian} and \ref{eq:trasmittance}, after having dropped constant phase factors and grouped together constant amplitude terms:

\begin{multline}
    U(x,y,z) = U_0\exp\bigg[i\kappa ln_m\cos(\omega_mt)\sin(k_mx)\bigg]\times \\
    \times\exp \left[ \frac{-r^2}{w^2(z)}\right]
    \exp \left[ -i\kappa \frac{r^2}{2R(z)}\right].
    \label{eq:U0}
\end{multline}
It is very useful to rewrite the first exponential function of the above equation using the Bessel generating function

\begin{equation}
    \exp\left[(\chi/2)(\tau-1/\tau)\right]=\sum_{q=-\infty}^{+\infty}J_q(\chi)\tau^q.
\end{equation}
Using the following identifications:

\begin{subequations}
\begin{align}
    \chi/2&=\kappa ln_m\cos(\omega_mt)/2 \\
    \tau-1/\tau&=e^{ik_mx}-e^{-ik_mx}
\end{align}
\end{subequations}
we obtain

\begin{multline}
    \exp\bigg[i\kappa ln_m\cos(\omega_mt)\sin(k_mx)\bigg]=\\
    =\sum_{q=-\infty}^{+\infty}J_q\big[n_m\cos(\omega_mt)\big]\exp(ik_mqx).
\end{multline}
We can now calculate the propagation of the field described by equation \ref{eq:U0} through a converging lens. Assuming that the focal length of this lens is long enough, we can apply the slowly varying envelope approximation. From the scalar diffraction theory \cite{Goodman2017}, the field at the lens focal plane is proportional to the Fourier transform of the field after the AOF cavity, and can be written as the product of three terms:
\begin{subequations}
\begin{align}
    &\mathscr{F}\left\{\sum_{q=-\infty}^{+\infty}J_q\big[\kappa ln_m\cos(\omega_mt)\big]\exp(ik_mqx)\right\}= \nonumber\\
    &=\sum_{q=-\infty}^{+\infty}J_q\big[\kappa ln_m\cos(\omega_mt)\big]\delta\left(\nu_x-\frac{qk_m}{2\pi}\right)\delta(\nu_y) \\
    &\mathscr{F}\left\{\exp \left[ \frac{-r^2}{w^2(z)}\right]\right\}={\pi w^2(z)}\exp\big[ -\pi^2w^2(z)\nu^2 \big] \\
    &\mathscr{F}\left\{\exp \left[-i\kappa \frac{r^2}{2R(z)}\right]\right\}={\dfrac{2\pi R(z)}{ik}}\exp\left[ -\pi^2\frac{2R(z)}{i\kappa}\nu^2 \right]
\end{align}
\end{subequations}
where $\nu_x$ and $\nu_y$ correspond to the spatial frequencies along the $x$ and $y$ axis.
Using the convolution property, we can write the Fourier transform of the product of the first two terms as

\begin{multline}
    \sqrt{\pi w^2(z)}\sum_{q=-\infty}^{+\infty}J_q\big[\kappa ln_m\cos(\omega_mt)\big]\times \\
    \times\exp\left[ -\pi^2w^2(z)\left[\left(\nu_x-\frac{qk_m}{2\pi}\right)^2+\nu_y^2\right] \right].
\end{multline}
 
\noindent The above equation corresponds to a sum of Gaussian functions. Because the convolution of two normalized Gaussian functions is still a normalized Gaussian function with mean $\vect{\mu}=\vect{\mu}_1+\vect{\mu}_2$ and variance $\sigma^2=\sigma_1^2+\sigma_2^2$, we can write the total Fourier transform as
\begin{align}
    \mathscr{F}\left\{U(x,y,t)\right\} = &U_0\dfrac{2\pi}{\dfrac{2}{w^2}+\dfrac{ik}{R}}\times \nonumber \\
    &\times\sum_{q=-\infty}^{+\infty}J_q\big[\kappa ln_m\cos(\omega_mt)\big] \times \nonumber \\
    &\times\exp\left[-\frac{\left(\nu_x-\dfrac{qk_m}{2\pi}\right)^2+\nu_y^2}{2\sigma_R^2} \right]\times \nonumber \\
    &\times\exp\left[i\frac{\left(\nu_x-\dfrac{qk_m}{2\pi}\right)^2+\nu_y^2}{2\sigma_I^2}\right],
\end{align}
\noindent where 
\begin{align}
    \frac{1}{\sigma_R^2}&=\frac{(2\pi^2w^2)^{-1}}{(2\pi^2w^2)^{-2}+\left(\dfrac{\kappa}{4\pi^2R}\right)} \\
    \frac{1}{\sigma_I^2}&=\frac{\left(\dfrac{\kappa}{4\pi^2R}\right)}{(2\pi^2w^2)^{-2}+\left(\dfrac{\kappa}{4\pi^2R}\right)}.
\end{align}

\noindent The intensity can be calculated as the square modulus of the above equation. Suppose now that $\omega_m$ is big enough to prevent the overlap of neighbouring Gaussian functions. Under this hypothesis we can neglect the cross-product of the terms of the sum. By identifying $(\nu_x,\nu_y)$ as $\frac{1}{\lambda f}(u,v)$, where $u$ and $v$ are the spatial coordinates at the focal plane of the lens, we can write the intensity function as

\begin{align}
    &I(u,v)=I_0\sum_{q=-\infty}^{+\infty}J^2_q\big[\kappa ln_m\cos(\omega_mt)\big]\times\\
    &\times\exp\left[ -\frac{2\pi^2w^2}{1+\dfrac{\kappa\pi^2w^4}{R}}\left[\left(\frac{u}{\lambda f}-\frac{qk_m}{2\pi}\right)^2+\left(\frac{v}{\lambda f}\right)^2\right] \right], \nonumber
\end{align}
where $I_0$ is the square modulus of all the constant prefactors divided by the impedance of free space. Therefore, the separation between peaks is 

\begin{equation}
    \mu_{q+1}-\mu_q=\frac{\omega_m}{2\pi c_s}\lambda f
\end{equation}
and the diffraction efficiency, defined as the peak intensity divided by $I_0$, is

\begin{equation}
    \eta_q(t)=J^2_q\big[\kappa ln_m\cos(\omega_mt)\big]
    \label{eq:eta_q_app}
\end{equation}
By following the same procedure described above, it is possible to obtain a generalized form of the multi-focus distribution for two orthogonal couples of piezoelectric plates:

\begin{align}
    &I(u,v)=I_0 \times \\
    &\times \sum_{q=-\infty}^{+\infty}
    \eta_q(t) \exp\left[ -\frac{2\pi^2w^2}{1+\dfrac{\kappa\pi^2w^4}{R}}\left(\frac{u}{\lambda f}-\frac{qk_m}{2\pi}\right)^2\right] \times \nonumber \\
    &\times \sum_{p=-\infty}^{+\infty}\eta_p(t)
    \exp\left[ -\frac{2\pi^2w^2}{1+\dfrac{\kappa\pi^2w^4}{R}}\left(\frac{v}{\lambda f}-\frac{pk_l}{2\pi}\right)^2\right]. \nonumber
\end{align}

\subsection{Measurement of the change in refractive index}
The good agreement between experimental results and model allowed us to calculate the change in refractive index induced in the cavity. From the temporal characterization of the AOF device, we obtained the intensity of each diffracted order as a function of time (Figure \hyperref[fig:eta-fit]{\ref{fig:eta-fit}b}) in the main text). By using equation \ref{eq:eta_q_app} to fit these data, we measured a value of $a=\kappa ln_m$ of $\SI{3.49(6)}{}$ for $V_{pp}=\SI{20}{V}$ and $\nu=\SI{1.488}{MHz}$. Knowing that $\kappa=\SI{1.412e7}{m^{-1}}$ (i.e. the wave number of the blue laser) and $L=\SI{16(1)}{mm}$ we obtained the maximum variation of refraction index of the liquid in the AOF cavity to be $n_m=\frac{a}{L\kappa}=\SI{1.5(1)e-5}{}$.

\vspace{1cm}
\bibliography{references.bib}

\end{document}